\documentclass[preprint,aps,nofootinbib]{revtex4}
\usepackage{amsfonts,amsmath,amssymb,amsthm}
\usepackage{latexsym}
\usepackage{bbm,bm}
\usepackage{graphicx}

\newcommand{\ket}[1]{\lvert #1 \rangle}

\newcommand{\beq}{\begin{equation}}
\newcommand{\eeq}{\end{equation}}
\newcommand{\beqs}{\begin{eqnarray}}
\newcommand{\eeqs}{\end{eqnarray}}

\begin{document}

\title{Quantum Illumination with three-mode Gaussian State}

\author{Eylee Jung$^1$, and DaeKil Park$^{1,2}$\footnote{dkpark@kyungnam.ac.kr} }

\affiliation{$^1$Department of Electronic Engineering, Kyungnam University, Changwon
                631-701, Korea    \\
            $^2$Department of Physics, Kyungnam University, Changwon
                  631-701, Korea    
                      }

\begin{abstract}
The quantum illumination is examined by making use of the three-mode maximally entangled Gaussian state, which involves one signal and two idler beams. 
It is shown that the quantum Bhattacharyya bound between $\rho$ (state for target absence) and $\sigma$ (state for target presence) is less than the previous result derived by two-mode Gaussian state 
when $N_S$, average photon number per signal, is less than $0.295$. This indicates that the quantum illumination with three-mode Gaussian state gives less error probability compared to that with two-mode Gaussian state
when $N_S < 0.295$. 
\end{abstract}
\maketitle

\section{Introduction}
As IC (integrated circuit) becomes smaller and smaller in modern classical technology, the effect of quantum mechanics becomes prominent more and more. As a result, quantum technology (technology 
based on quantum mechanics and quantum information theories\cite{text}) becomes important more and more recently. The representative constructed by quantum technology
is a quantum computer\cite{supremacy-1}, which was realized recently by making use of superconducting qubits. 
In quantum information processing quantum entanglement\cite{text,schrodinger-35,horodecki09} plays an important role as a physical resource. 
It is used in various quantum information processing, such as  quantum teleportation\cite{teleportation,Luo2019},
superdense coding\cite{superdense}, quantum cloning\cite{clon}, quantum cryptography\cite{cryptography,cryptography2}, quantum
metrology\cite{metro17}, and quantum computer\cite{supremacy-1,qcreview,computer}. 

Few years ago the entanglement-assisted target detection protocol called the quantum illumination\cite{lloyd08,tan08} and its experimental realization\cite{guha09,lopaeva13,barzanjeh15,zhang15,zhuang17} were explored. 
Quantum illumination is a protocol which takes advantage of quantum entanglement to detect a low reflective object 
embedded in a noisy thermal bath. The typical quantum illumination can be described as follows. The transmitter generates two entangled photons called signal (S) and idler (I) modes. The S-mode photon is used to interrogate 
the unknown object hidden in the background. After receiving photon from a target region, joint quantum measurement for returned light and I-mode photon is performed to decide absence or presence of target.
The most surprising result of the quantum illumination is the fact that the error probability for the detection is drastically lowed even if the initial entanglement between S and I modes disappears 
due to decoherence. To show more concretely let us briefly review Ref. \cite{tan08}, where the quantum illumination was explored with two-mode Gaussian state. In the paper the initial state is chosen as a maximally entangled state  as 
a form
\begin{equation}
\label{two-signal-state}
\ket{\psi}_{SI} = \sum_{n=0}^{\infty} \sqrt{\frac{N_S^n}{(1 + N_S)^{n+1}}} \ket{n}_S \ket{n}_I,
\end{equation}
where $N_S$ is the average photon number per signal mode. This is a zero-mean Gaussian state whose covariance matrix is 
\begin{eqnarray}
\label{two-signal-state-2}
\Lambda_{SI}^{(2)} = \left(     \begin{array}{cccc}
                           S  &  0  &  C_q  &  0            \\
                           0  &  S  &  0  &  -C_q           \\
                           C_q  &  0  &  S  &  0            \\
                           0  &  -C_q  &  0  &  S
                                \end{array}               \right)
\end{eqnarray}
where $S \equiv 2 N_S + 1$ and $C_q = 2 \sqrt{N_S (1 + N_S)}$. Even though $\ket{\psi}_{SI}$ is maximally entangled, it was shown  that $\rho$ (state for target absence) and
$\sigma$ (state for target presence) are separable due to entanglement-breaking noise. 
The most important result of this paper is as follows. Let $N_B$ and $\kappa$ be average photon number of background thermal state and  reflectivity from a target respectively. 
Let us assume that background contribution is very strong ($N_B \gg 1$) compared to the signal $(N_B \gg N_S)$. We also assume that the reflectivity from a target is extremely small $(\kappa \ll 1)$.
Then,  the quantum Bhattacharyya (QB) bound between $\rho$ and $\sigma$ is shown to be
\begin{equation}
\label{QB-2}
P_{QB}^{(2)} \approx \frac{1}{2} \exp \left[ - \frac{M \kappa \gamma_2}{N_B} \right]
\end{equation}
where $M$ is a number of identical copies of $\rho$ and $\sigma$, and $\gamma_2$ is 
\begin{equation}
\label{QB-2-1}
\gamma_2 = \frac{N_S (1 + N_S) \left[1 + N_S - \sqrt{N_S (1 + N_S)} \right]}{1 + N_S + \sqrt{N_S (1 + N_S)}} 
= N_S \left[ 1 - 2 \sqrt{N_S} + {\cal O} (N_S) \right].
\end{equation}
The last equality in Eq. (\ref{QB-2-1}) is valid only when $N_S \ll 1$. This is important because the QB bound for the single-mode coherent-state is 
\begin{equation}
\label{QB-1}
P_{QB}^{(1)} \approx \frac{1}{2} \exp \left[ - \frac{M \kappa N_S}{4 N_B} \right].
\end{equation}
When $N_S \ll 1$, the difference of Eq. (\ref{QB-2}) from Eq. (\ref{QB-1}) is a missing of factor $4$ in the exponent. 
This implies that the quantum illumination with two-mode Gaussian state gives much less error probability compared to that with the classical coherent state.
The quantum illumination with two-mode Gaussian state was extended to the asymmetric Gaussian hypothesis testing\cite{asymmetry-1,asymmetry-2}. Also, the quantum illumination with non-Gaussian initial state generated by photon subtraction and addition 
was also discussed \cite{zhang14,fan18}.

\begin{figure}[ht!]
\begin{center}
\includegraphics[height=6.0cm]{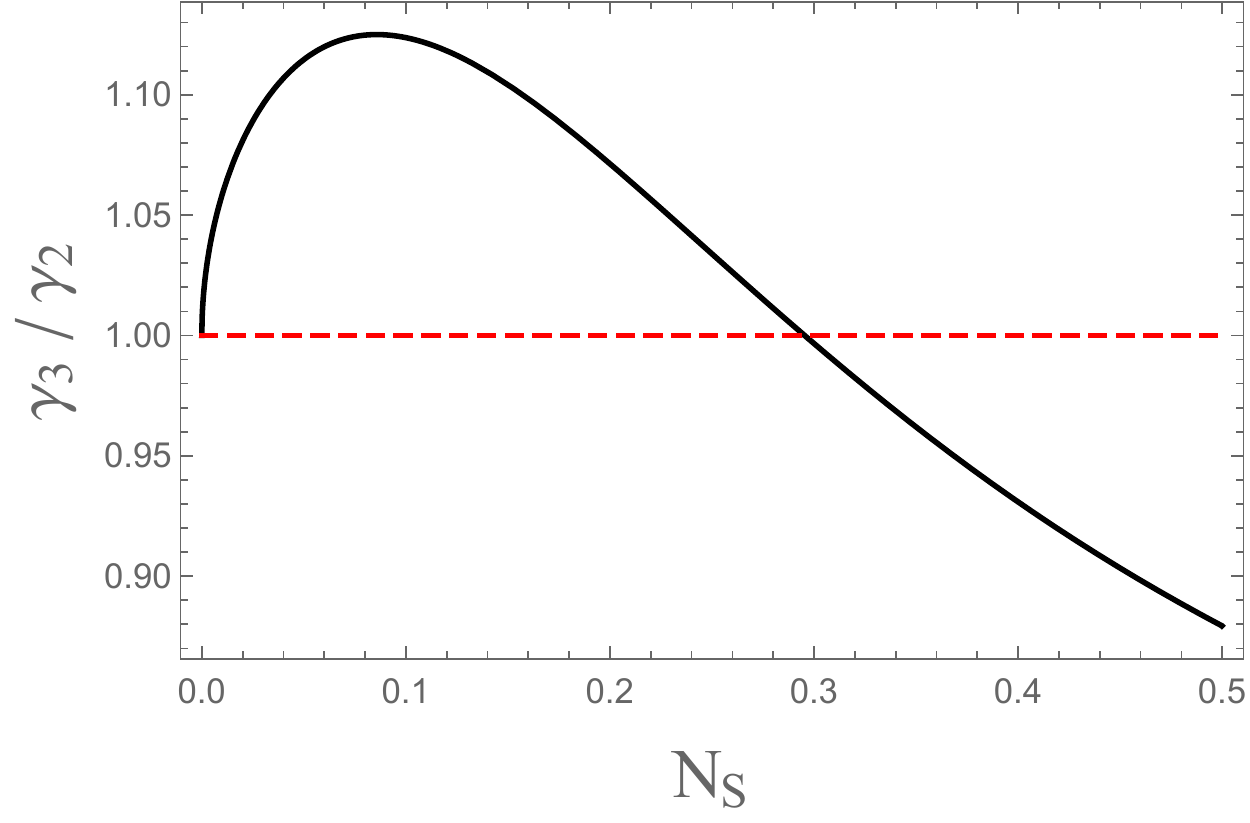} 

\caption[fig1]{(Color online) The $N_S$-dependence of the ratio $\gamma_3 / \gamma_2$
 }
\end{center}
\end{figure}

As emphasized above, the entanglement's benefit survives even though there is no entanglement in $\rho$ and $\sigma$ due to entanglement-breaking noise. 
Then, it is natural to ask how entanglement's benefit is enhanced when partial entanglement of the initial state survives in  $\rho$ and $\sigma$. 
In order to examine this issue we consider the same physical situation in this paper by making use of the three-mode Gaussian state (one signal ($S$) and two idler modes ($I_1$, $I_2$)). 
Thus, there are three biparties $(S, I_1 + I_2)$, $(I_1, S + I_2)$, $(I_2, S + I_1)$. 
As we will show in the following, the states $\rho$ and $\sigma$ derived from the three-mode maximally entangled state still have zero bipartite entanglement for first biparty while non-zero for the remaining parties. In this reason the noise discussed in our quantum illumination scheme is not completely 
entanglement-breaking, because it survives between two idler beams.
Experimental construction of the three-mode states was recently discussed in Ref. \cite{chang20}. 

\section{Main Result}
The final and main result of this paper is that the QB bound for our case becomes
\begin{equation}
\label{QB-3}
P_{QB}^{(3)} \approx \frac{1}{2} \exp \left[ - \frac{M \kappa \gamma_3}{N_B} \right]
\end{equation}
where 
\begin{equation}
\label{QB-3-1}
\gamma_3 \approx N_S \left[1 -  \sqrt{2 N_S} + {\cal O} (N_S) \right]
\end{equation}
when $N_S \ll 1$. 
Thus, this QB bounds is slightly different from Eq. (\ref{QB-2}) in the second order of exponent. As a result, $P_{QB}^{(3)}$ is slightly less than $P_{QB}^{(2)}$. 
With relaxing the criterion $N_S \ll 1$, one can compute the ratio $\gamma_3 / \gamma_2$, which is plotted in Fig. 1. Fig. 1 indicates that $\gamma_3$ is larger 
than $\gamma_2$ in the region $0 \leq N_S \leq 0.295$. This indicates that the quantum illumination with three-mode Gaussian state gives less error probability compared to that with two-mode Gaussian state
when $N_S < 0.295$.

\section{Theory}
In order to derive Eq. (\ref{QB-3}) we start with a state of signal and two idler modes. We assume that this state is a zero-mean three-mode Gaussian state whose covariance matrix is 
\begin{eqnarray}
\label{signal-state}
\Lambda_{S I_1 I_2}^{(3)} = \left( \begin{array}{cccccc}
                          S  &  0  &  C  &  0  &  C  &  0         \\
                          0  &  S  &  0  &  -C  &  0  &  -C       \\
                          C  &  0  &  S  &  0  &  C  &  0         \\
                          0  &  -C  &  0  &  S  &  0  &  -C       \\
                          C  &  0  &  C  &  0  &  S  &  0         \\
                          0  &  -C  &  0  &  -C  &  0  &  S
                               \end{array}                \right)
\end{eqnarray}
where $S = 2 N_S + 1$. The subscript ``$S I_1 I_2$'' stands for signal, idler $1$, and idler $2$ respectively. Given the diagonal elements of $\Lambda_{S I_1 I_2}^{(3)}$, 
quantum mechanics allows that the non-zero off-diagonal elements, $C$, should obey $0 \leq C \leq C_q^{(3)}$, where $\left(C_q^{(3)} \right)^2$ is a root 
of $4 x^3 - 9 S^2 x^2 + 6 S^4 x - (S^6 - 1) = 0$. This cubic equation gives unique real root in the range $0 < x < S^2 / 2$. 
Since any cubic equation can be analytically solved in principle, $C_q^{(3)}$ can be completely 
determined. Although it is hard to express $C_q^{(3)}$ explicitly due to its lengthy expression, one can show that it has following limits:
\begin{eqnarray}
\label{limiting}
C_q^{(3)} = \left\{   \begin{array}{ll}
        \sqrt{2 N_S} \left[ 1 - \frac{2}{3} N_S^2 + \frac{4}{3} N_S^3 + \cdots  \right] & \hspace{2.0cm}  N_S \ll 1    \\
        N_S\left[ 1 + \frac{1}{2 N_S} - \frac{1}{72} \frac{1}{N_S^6} + \cdots \right].   & \hspace{2.0cm}  N_S \gg 1
                      \end{array}         \right.
\end{eqnarray}
When $C = C_q^{(3)}$, the composite system becomes pure and maximally entangled state between the signal and idler modes. When $0 \leq C \leq C_c^{(3)}$, where 
$\left(C_c^{(3)} \right)^2 = \left[ (2 + 5 N_S + 5 N_S^2) - \sqrt{(1 + 3 N_S) (2 + 3 N_S) (2 + N_S + N_S^2)} \right] / 2$, the composite state becomes separable.

Let $\rho$ and $\sigma$ be quantum states for target absence and target presence respectively. Both are the zero-mean Gaussian states. Since, for $\rho$, the annihilation operator 
for the return from the target region should be $\hat{a}_R = \hat{a}_B$, where $\hat{a}_B$ is the annihilation operator for a thermal state with average photon number $N_B$, its covariance matrix 
can be written in a form:
\begin{eqnarray}
\label{rho-state}
\Lambda_{\rho}  = \left( \begin{array}{cccccc}
                          B  &  0  &  0  &  0  &  0  &  0         \\
                          0  &  B  &  0  &  0  &  0  &  0       \\
                          0  &  0  &  S  &  0  &  C  &  0         \\
                          0  &  0  &  0  &  S  &  0  &  -C       \\
                          0  &  0  &  C  &  0  &  S  &  0         \\
                          0  &  0  &  0  &  -C  &  0  &  S
                               \end{array}                \right)
\end{eqnarray}
where $B = 2 N_B + 1$. It can be written as $\Lambda_{\rho} = S_{\rho}  \left(\bigoplus_{j = 1}^3 \alpha_j \openone_2 \right) S_{\rho}^T$, where the symplectic eigenvalues are 
\begin{equation}
\label{symplectic-1}
\alpha_1 = B  \hspace{1.0cm}  \alpha_2 = \alpha_3 = \sqrt{S^2 - C^2}.
\end{equation}
The symplecic transformation $S_{\rho}$ is 
\begin{eqnarray}
\label{symplectic-2}
S_{\rho} = \left( \begin{array}{ccc}
              \openone  &  0  &  0                    \\
              0  &  Z_1  &  -Z_2                      \\
              0  &  Z_1  &  Z_2
                  \end{array}       \right)
\end{eqnarray}
where $Z_1 = \frac{1}{\sqrt{2}} \mbox{diag} (1 / z, z)$ and $Z_2 = \frac{1}{\sqrt{2}} \mbox{diag} (z, 1 / z)$ with $z = \left[ (S - C) / (S + C) \right]^{1/4}$.
When $C = C_q^{(3)}$, the state $\rho$ has non-zero logarithmic negativity ${\cal E}_N = - \log_2 \left(S - C_q^{(3)} \right)$ in the bipartition $I_1 + (S, I_2)$ and $I_2 + (S, I_1)$. However, it is zero 
in the party $S + (I_1, I_2)$. 

For $\sigma$ the return-mode's annihilation operator would be $\hat{a}_R = \sqrt{\kappa} \hat{a}_S + \sqrt{1 - \kappa} \hat{a}_B$, 
where $\hat{a}_B$ is an annihilation operator for a thermal state with average photon number $N_B / (1 - \kappa)$. We assume $\kappa \ll 1$.
Combining all of the facts, one can deduce that the covariance matrix for $\sigma$ is 
\begin{eqnarray}
\label{sigma-state}
\Lambda_{\sigma} = \left( \begin{array}{cccccc}
                          A  &  0  &  \sqrt{\kappa} C  &  0  & \sqrt{\kappa} C  &  0         \\
                          0  &  A  &  0  &  - \sqrt{\kappa}C  &  0  &  - \sqrt{\kappa}C       \\
                          \sqrt{\kappa} C  &  0  &  S  &  0  &  C  &  0         \\
                          0  &  - \sqrt{\kappa} C  &  0  &  S  &  0  &  -C       \\
                          \sqrt{\kappa} C  &  0  &  C  &  0  &  S  &  0         \\
                          0  &  - \sqrt{\kappa} C  &  0  &  -C  &  0  &  S
                               \end{array}                \right)
\end{eqnarray}
where $A = 2 \kappa N_S + B$. The symplectic eigenvalues of $\Lambda_{\sigma}$ are 
\begin{eqnarray}
\label{symplectic-3}
&&\hspace{6.0cm}  \beta_1 = \sqrt{S^2 - C^2}   \\   \nonumber
&& \beta_2 = \sqrt{\frac{A^2 + S^2 - (1 + 4 \kappa) C^2 + \xi}{2}} \equiv \beta_+   \hspace{.5cm}  \beta_3 = \sqrt{\frac{A^2 + S^2 - (1 + 4 \kappa) C^2 - \xi}{2}} \equiv \beta_{-}
\end{eqnarray}
where $\xi = \sqrt{(A^2 - S^2 + C^2)^2 - 8 \kappa C^2 (A - S + C) (A - S - C)}$. Then, it is straightforward to show  $\Lambda_{\sigma} = S_{\sigma}  \left(\bigoplus_{j = 1}^3 \beta_j \openone_2 \right) S_{\sigma}^T$, where
\begin{eqnarray}
\label{symplectic-4}
S_{\sigma} = \left( \begin{array}{ccc}
              0  &  X_1  &  X_2                    \\
              Z_2  & Y_1  & Y_2                      \\
              -Z_2  &  Y_1  &  Y_2
                  \end{array}       \right).
\end{eqnarray}
The $2 \times 2$ matrices $X_j$ and $Y_j$ are $X_1 = \mbox{diag} (x_+, y_+)$, $X_2 = \mbox{diag} (y_-, x_-)$, $Y_1 = \mbox{diag} (u_+, v_+)$, and $Y_2 = \mbox{diag} (v_-, u_-)$ where
\begin{eqnarray}
\label{sym-11}
&&  x_{\pm} = \pm \frac{1}{2} \sqrt{ \frac{\mu_{1, \pm} \mu_{2, \pm}}{(A - S \mp C) \xi \beta_{\pm}}}  \hspace{1.0cm}
 y_{\pm} =  \sqrt{ \frac{(A - S \mp C)  \mu_{2, \pm} \beta_{\pm}}{\mu_{1, \pm} \xi }}                            \\    \nonumber
&&  u_{\pm} = \sqrt{ \frac{\mu_{1, \pm} \mu_{2, \mp}}{8 (A - S \pm C) \xi \beta_{\pm}}}    \hspace{1.0cm}
v_{\pm} = \mp   \sqrt{ \frac{(A - S \pm C)  \mu_{2, \mp} \beta_{\pm}}{2 \mu_{1, \pm} \xi }}.
\end{eqnarray}
In Eq. (\ref{sym-11}) $\mu_{1, \pm}$ and $\mu_{2, \pm}$ are given by $\mu_{1,\pm} = (\xi - 2 A C) \pm \left[(A - S)^2 - C^2 \right]$ and $\mu_{2, \pm} = A^2 - S^2 + C^2 \pm \xi$. One can show explicitly 
\begin{eqnarray}
\label{useful}
&&   \mu_{2,+} - \mu_{2,-} = 2 \xi \hspace{1.0cm} \mu_{2,+} \mu_{2,-} = 8 \kappa C^2 (A - S + C) (A - S - C)            \\   \nonumber
&& \mu_{1,+} \mu_{2,+} = 2 (A - S - C) \left[ A \mu_{2,+} - 4 \kappa C^2 (A - S + C) \right]                         \\  \nonumber
&&\mu_{1,-} \mu_{2,-} = -2 (A - S + C) \left[ A \mu_{2,-} - 4 \kappa C^2 (A - S - C) \right].
\end{eqnarray}
These relations are frequently used in the calculation of the QB bound. 
The state $\sigma$ has non-zero logarithmic negativity in the bipartition $I_1 + (S, I_2)$ and $I_2 + (S, I_1)$, whose explicit expression is too complicated to express it explicitly.
Like a $\rho$, however, it is zero in the party $S + (I_1, I_2)$

In order to accomplish the quantum illumination processing, we should choose the one of the null hypothesis $H_0$ (target absent) or the alternative hypothesis $H_1$ (target present). Then, the average error probability is 
\begin{equation}
\label{error-1}
P_{E} = P(H_0) P (H_1 | H_0) + P(H_1) P(H_0 | H_1)
\end{equation}
where $P(H_0)$ and $P(H_1)$ are the prior probabilities associated with the two hypotheses. We assume $P(H_0) = P(H_1) = 1 / 2$ for simplicity.
Therefore, the minimization of $P_E$ naturally requires the optimal discrimination of $\rho$ and $\sigma$. 
If we have $M$ identical copies of $\rho$ and $\sigma$, the optimal discrimination scheme presented in Ref. \cite{sacchi05-1,sacchi05-2} yields the minimal error probability $P_E^{min}$ in a form
\begin{equation}
\label{error-2}
P_E^{min} = \frac{1}{2} \left[ 1 - \frac{1}{2} || \rho^{\otimes M} - \sigma^{\otimes M} ||_1 \right]
\end{equation}
where $||A||_1 = \mbox{Tr} \sqrt{A^{\dagger} A}$ denotes the trace norm of $A$. However, the computation of the trace norm in Eq. (\ref{error-2}) seems to be highly tedious for large $M$. Also, it is difficult to imagine the large $M$ behavior of 
the minimal error probability from Eq. (\ref{error-2}). In order to overcome these difficulties the quantum Chernoff bound was considered\cite{chernoff-1,chernoff-2}. The quantum Chernoff bound $P_{QC}$ of $\rho$ and $\sigma$ is defined as 
\begin{equation}
\label{chernoff-1}
P_{QC} = \frac{1}{2} \left(\min_{s \in [0,1]} \mbox{Tr} \left[\rho^s \sigma^{1-s} \right]\right)^M.
\end{equation}
This is a tight upper bound of $P_{E}^{min}$, i.e., $P_{E}^{min} \leq P_{QC}$. This bound was analytically computed in several simple quantum systems \cite{chernoff-2}. However, the computation of the 
optimal $s$, which minimizes $\mbox{Tr} \left[\rho^s \sigma^{1-s} \right]$, is in general highly tedious. 

The Chernoff bound for general $n$-mode Gaussian states $\tilde{\rho}$ and $\tilde{\sigma}$ can be expressed as follows.
First we define 
\begin{equation}
\label{func-1}
\Lambda_p (x) \equiv \frac{(x + 1)^p + (x - 1)^p}{(x + 1)^p - (x - 1)^p}   \hspace{1.0cm} G_p (x) \equiv \frac{2^p}{(x + 1)^p - (x - 1)^p}.
\end{equation}
Let the mean displacement vector of $\tilde{\rho}$ and $\tilde{\sigma}$ be $\bar{x}_{\rho}$ and  $\bar{x}_{\sigma}$, and the corresponding covariance matrices are 
$\tilde{\Lambda}_{\rho} = S_{\rho}  \left(\bigoplus_{j = 1}^n \alpha_j \openone_2 \right) S_{\rho}^T$ and $\tilde{\Lambda}_{\sigma} = S_{\sigma}  \left(\bigoplus_{j = 1}^n \beta_j \openone_2 \right) S_{\sigma}^T$.
Now, we define $V_{\rho} (s) = S_{\rho}  \left(\bigoplus_{j = 1}^n \Lambda_s \left(\alpha_j\right) \openone_2 \right) S_{\rho}^T$ and 
$V_{\sigma} (s) =  S_{\sigma}  \left(\bigoplus_{j = 1}^n \Lambda_{s} \left(\beta_j\right) \openone_2 \right) S_{\sigma}^T$.
Then, the Chernoff bound for $\tilde{\rho}$ and $\tilde{\sigma}$ \cite{computable-1} is expressed as 
\begin{equation}
\label{gaus-cher-1}
P_{QC} = \frac{1}{2} \left[\min_{s \in [0,1]} Q_s \right]^M
\end{equation}
where 
\begin{equation}
\label{gaus-cher-2}
Q_s = \bar{Q}_s \exp \left[ -d^T \left( V_{\rho} (s) + V_{\sigma} (1 - s) \right)^{-1} d \right].
\end{equation}
In Eq. (\ref{gaus-cher-2}) $d = \bar{x}_{\rho} - \bar{x}_{\sigma}$ and 
\begin{equation}
\label{gaus-cher-3}
\bar{Q}_s = \frac{2^n \prod_{k=1}^n G_s (\alpha_k) G_{1 - s} (\beta_k)}{\sqrt{\det \left[V_{\rho} (s) + V_{\sigma} (1 - s) \right]}}.
\end{equation}

Now, let us return to our case. 
Since $d = 0$ and $n = 3$ for our case, the quantum Chernoff bound for $\rho$ and $\sigma$ can be written as 
\begin{equation}
\label{gaus-cher-4} 
P_{QC} = \frac{1}{2} \left( \min_{s \in [0,1]} \bar{Q}_s \right)^M.
\end{equation}
In Eq. (\ref{gaus-cher-4}) it is highly tedious and complicated process to find the optimal $s$, which minimizes $\bar{Q}_s$. 
Therefore, the QB bound $P_{QB}$ can be considered, where $s = 1/2$ is chosen instead of the optimal $s$. 
In this reason, $P_{QB}$ is always larger than $P_{QC}$ if the optimal $s$ is not $1/2$. 
The QB bound for our case is 
\begin{equation}
\label{gaus-cher-5}
P_{QB} = \frac{1}{2} \left(  \bar{Q}_{s=1/2} \right)^M
\end{equation}
where 
\begin{equation}
\label{gaus-cher-6}
 \bar{Q}_{s=1/2} = \frac{2^3 \prod_{k=1}^3 G_{1/2} (\alpha_k) G_{1/2} (\beta_k)}{\sqrt{\det \left[V_{\rho} \left( \frac{1}{2} \right) + V_{\sigma} \left( \frac{1}{2} \right) \right]}}.
\end{equation}

Now let us assume again $N_B \gg 1$, $N_B \gg N_S$, and $\kappa \ll 1$.
Then, it is possible to show 
\begin{eqnarray}
\label{gaus-qb-1}
&&  2^3 \prod_{k=1}^3 G_{1/2} (\alpha_k) G_{1/2} (\beta_k)                   \\   \nonumber
&&= \frac{128 N_B}{\left( \sqrt{\nu + 1} - \sqrt{\nu - 1} \right)^4} \left[ 1 + \frac{1}{2 N_B} \left\{ 1 + \kappa N_S - \frac{\kappa C^2 S}{\nu \sqrt{\nu^2 - 1}} \right\} + {\cal O} (N_B^{-2})  \right]  \\    \nonumber
&& \sqrt{\det \left[V_{\rho} \left( \frac{1}{2} \right) + V_{\sigma} \left( \frac{1}{2} \right) \right]}  =  \frac{128 N_B}{\left( \sqrt{\nu + 1} - \sqrt{\nu - 1} \right)^4}               \\     \nonumber
&& \times \left[ 1 + \frac{1}{2 N_B} \left\{ 1 + \kappa N_S + \kappa C^2 S  \left(1 - \frac{\sqrt{\nu^2 - 1}}{\nu} - \frac{1}{\nu \sqrt{\nu^2 - 1}} \right) \right\} + {\cal O} (N_B^{-2})  \right]
\end{eqnarray}
where $\nu = \sqrt{S^2 - C^2}$. Inserting Eq. (\ref{gaus-qb-1}) into Eq. (\ref{gaus-cher-5}) and putting $C = C_q^{(3)}$, one can derive the main result (\ref{QB-3}) where $\gamma_3$ is given by 
\begin{equation}
\label{QB-3-2}
\gamma_3 = \frac{1}{2} \left( C_q^{(3)} \right)^2 S \left(1 - \frac{\sqrt{\nu^2 - 1}}{\nu} \right) = N_S \left[ 1 - \sqrt{2 N_S} + {\cal O} (N_S) \right]
\end{equation}
and $\nu = \sqrt{S^2 - \left( C_q^{(3)} \right)^2}$. The last equality of Eq. (\ref{QB-3-2}) holds for only $N_S \ll 1$.

\section{Discussion}

In this paper we explore the quantum illumination by making use of the three-mode Gaussian state. If $N_S \ll 1$, the resulting QB bound is slightly less than the previous result derived in Ref. \cite{tan08}. 
Furthermore, we show that same is true when $0 \leq N_S \leq 0.295$. We guess the improvement of entanglement advantage 
in this region compared to the two-mode Gaussian illumination is due to the residual entanglement between  the two idler beams. However, this 
does not explain why the entanglement advantage is reduced when $N_S > 0.295$.  
In order to understand this phenomenon more clearly, it seems to be necessary to examine the $N$-mode Gaussian illumination. 

One can examine the advantage of the three-mode Gaussian approach in the asymmetric Gaussian hypothesis testing. 
Also, it is of interest to examine the effect of the Gaussian operations such as squeezing operation in quantum Gaussian illumination scheme. 
We hope to revisit these issues in the future.

{\bf Acknowledgement}:
This work was supported by the National Research Foundation of Korea(NRF) grant funded by the Korea government(MSIT) (No. 2021R1A2C1094580).

\end{document}